\documentclass[reprint,aps,prb]{revtex4-1}
\usepackage{color,graphicx,calc,url}
\usepackage{dcolumn}
\usepackage{bm}
\usepackage{amssymb}
\usepackage{amsmath}
\usepackage{wasysym}

\begin{document}
\title{Static magnetic proximity effect in Pt/NiFe$_2$O$_4$ and Pt/Fe bilayers investigated by x-ray resonant magnetic reflectivity}
\author{T.~Kuschel$^1$, C.~Klewe$^1$, J.-M.~Schmalhorst$^1$, F.~Bertram$^2$, O. Kuschel$^3$, T.~Schemme$^3$, J.~Wollschl\"ager$^3$, S.~Francoual$^2$, J.~Strempfer$^2$, A.~Gupta$^4$, M.~Meinert$^1$, G.~G\"otz$^1$, D.~Meier$^1$, G.~Reiss$^1$\email{Electronic mail: tkuschel@physik.uni-bielefeld.de}}
\affiliation{$^1$Center for Spinelectronic Materials and Devices, Department of Physics, Bielefeld University, Universit\"atsstra\ss e 25, 33615 Bielefeld, Germany\\
$^2$ Deutsches Elektronen-Synchrotron DESY, Notkestra\ss e 85, 22607 Hamburg, Germany\\
$^3$ Fachbereich Physik, Universit\"at Osnabr\"uck, Barbarastra\ss e 7, 49069 Osnabr\"uck, Germany\\
$^4$ Center for Materials for Information Technology, University of Alabama, Tuscaloosa Alabama 35487, USA}

\date{\today}

\keywords{}

\begin{abstract}
The spin polarization of Pt in Pt/NiFe$_2$O$_4$ and Pt/Fe bilayers is studied by interface-sensitive x-ray resonant magnetic reflectivity to investigate static magnetic proximity effects. The asymmetry ratio of the reflectivity was measured at the Pt $L_3$ absorption edge using circular polarized x-rays for opposite directions of the magnetization at room temperature. The results of the $2\%$ asymmetry ratio for Pt/Fe bilayers are independent of the Pt thickness between $1.8$ and $20\,\textrm{nm}$. By comparison with ab initio calculations, the maximum magnetic moment per spin polarized Pt atom at the interface is determined to be $(0.6\pm0.1)\,\mu_{B}$ for Pt/Fe. For Pt/NiFe$_2$O$_4$ the asymmetry ratio drops below the sensitivity limit of $0.02\,\mu_{B}$ per Pt atom. Therefore, we conclude, that the longitudinal spin Seebeck effect recently observed in Pt/NiFe$_2$O$_4$ is not influenced by a proximity induced anomalous Nernst effect.
\end{abstract}

\maketitle

In spintronics \cite{Wolf:2001} and spin caloritronics \cite{Bauer:2012fq} pure spin currents can be generated in ferromagnetic insulators~(FMIs) by spin pumping \cite{Tserkovnyak:2002jk}, the spin Hall effect \cite{Hirsch:1999hn} and the spin Seebeck effect \cite{Uchida:2008cc}. Since these spin currents play an important role in spintronic applications, an understanding of the generation, manipulation and detection of spin currents is an important topic of research. A common spin current detection technique uses a nonferromagnetic metal~(NM) thin film grown on a ferromagnet (FM). The inverse spin Hall effect \cite{Saitoh:2006kk} converts the spin current into a transverse voltage in the NM. Pt is commonly used as NM due to its large spin Hall angle \cite{Hoffmann:2013el}, but has generated some controversy in the interpretation because of its closeness to the Stoner criterion, which can induce, e.g., Hall or Nernst effects due to the proximity to the FM \cite{Huang:2012ks}. 

For a quantitative evaluation of the spin Seebeck effect (thermal generation of spin currents) one has to exclude or separate various parasitic effects. It is reported \cite{Uchida:2008cc} that in transverse spin Seebeck experiments a spin current is generated perpendicular to the applied temperature gradient which is typically aligned in-plane. For ferromagnetic metals (FMMs) with magnetic anisotropy, the planar Nernst effect \cite{Ky:1966b} can contribute \cite{Avery:2012bj} due to the anisotropic magnetothermopower. Furthermore, out-of-plane temperature gradients due to heat flow into the surrounding area \cite{Schmid:2013ba} or through the electrical contacts \cite{Meier:2013fa} can induce an anomalous Nernst effect (ANE) \cite{Nernst:1886,Bosu:2011bw,Huang:2011io} or even an unintended longitudinal spin Seebeck effect as recently reported \cite{Meier2014}. 

The longitudinal spin Seebeck effect (LSSE) \cite{Uchida:2010jb} describes a spin current that is generated parallel to the temperature gradient, which is typically aligned out-of-plane to drive the parallel spin current directly into the NM material. For FMMs or semiconducting ferromagnets an ANE can also contribute to the measured voltage \cite{Meier:2013dz}. Furthermore, for NM materials close to the Stoner criterion a static magnetic proximity effect in the NM at the NM/FMI interface can lead to a proximity induced ANE \cite{Huang:2012ks}. If an in-plane temperature gradient is applied, a proximity induced planar Nernst effect should also be considered \cite{Meier2014}.


Magnetic proximity effects in Pt have been investigated intensively for multilayers with Pt adjacent to FMMs like Fe\cite{Antel:1999}, Co\cite{Schuetz:1990sd,Wilhelm:2003di}, and Ni \cite{Wilhelm:2000} using x-ray magnetic circular dichroism (XMCD). However, the investigation of Pt/FMI is new and needs interface-sensitive techniques. So far, XMCD in fluorescent mode was used for Pt/FMI to investigate the magnetic proximity in Pt attached to the FMI Y$_3$Fe$_5$O$_{12}$ (YIG). Gepr\"ags et al. \cite{Gepraegs:2012be,Geprags:2013} did not observe any evidence for a spin polarization in $3\,\textrm{nm}$ and $1.6\,\textrm{nm}$ thick Pt on YIG, while Lu et al. \cite{Lu:2013eq} detected an average magnetic moment of $0.054\,\mu_{\textrm{B}}$ at $300\,\textrm{K}$ and $0.076\,\mu_{\textrm{B}}$ at $20\,\textrm{K}$ in $1.5\,\textrm{nm}$ Pt on YIG.

While XMCD only gives the mean polarization of the Pt film, x-ray resonant magnetic reflectivity (XRMR) \cite{Macke:2014em} is directly sensitive to the spin polarization at the interface due to the interference of reflected light from the Pt surface and the Pt/FM interface. The magnetic dichroism for circularly polarized light reflected by the spin polarized interface results in a slightly different x-ray reflectivity (XRR) for varying magnetization directions due to a change of the optical constants of the spin polarized material. In particular, in magnetic materials the refractive index $n=1-\delta+i\beta$ with dispersion and absorption coefficients $\delta$ and $\beta$ changes with $\pm\Delta\delta$ and $\pm\Delta\beta$ for different magnetization directions ($\pm$). The sensitivity of XRMR to the interface spin polarization additionally allows us to evaluate magnetooptic depth profiles of $\Delta\delta$ and $\Delta\beta$ which cannot be provided by XMCD. Furthermore, an oscillating asymmetry ratio extending over a wide range of scattering vectors can be identified much more reliably than one weak peak produced by XMCD. Thus, the limit is extended to much smaller values of spin polarization.

In recent LSSE and ANE \cite{Meier:2013dz} and spin Hall magnetoresistance studies \cite{Althammer:2013ck} on Pt/NiFe$_2$O$_4$ (NFO) bilayers, we separated LSSE and ANE by measurements with and without Pt. The electrically semiconducting NFO allows a comparison of the temperature-dependent conductivity with spin Seebeck coefficients. In this letter, we focus on XRMR measurements of Pt/NFO to identify or exclude additional proximity induced ANEs that have not been taken into account before.

We prepared NFO films of up to $1\,\mu\textrm{m}$ thickness on $10\times5\,\textrm{mm}^2$ MgAl$_2$O$_4$(001) (MAO) substrates by direct liquid injection chemical vapor deposition \cite{Meier:2013dz, Li:2011}. Furthermore, we deposited Fe on MAO as reference sample with Pt on top. Both Pt and Fe were deposited by dc magnetron sputtering in a $1.5\times 10^{-3}\,\textrm{mbar}$ Ar atmosphere.

XRMR was measured at room temperature at the resonant scattering and diffraction beamline P09 of the third-generation synchrotron PETRA III at DESY (Hamburg, Germany) \cite{Strempfer:2013th}. The sample and magnet system were mounted on the six-circle diffractometer in the P09 first experimental hutch (P09-EH1) and the reflectivity curves were taken in a $\theta-2\theta$ scattering geometry. The external magnetic field was applied in the scattering plane and parallel to the sample surface using a four coils electromagnet constructed at Bielefeld University. Circularly polarized x-rays were generated by two diamond plates at the eight-wave plate condition mounted in series. The degree of polarization was $(99\pm1)\%$ for right and left circular polarization as determined from a polarization analysis with a Au(111) analyzer crystal. 

For detecting the Pt polarization, we used left circularly polarized x-rays with the photon energy of $11567.5\,\textrm{eV}$ (Pt $L_3$ absorption edge) and changed the magnetic field direction after having confirmed that right circularly polarized x-rays changes the sign of the XRMR asymmetry ratio. An external magnetic field of $\pm85\textrm{mT}$ was applied for each angle of incidence $\theta$ while the reflected intensity $I_{\pm}$ was detected. The dependence of the nonmagnetic reflectivity $I=\frac{I_++I_-}{2}$ on the scattering vector $q=\frac{4\,\pi}{\lambda}\,\sin\theta$ ($\lambda$ is the wavelength) was fitted using the recursive Parratt algorithm \cite{Parratt:1954fy} and a N\'evot-Croce roughness model \cite{Nevot:1980} using the analysis tool \textit{i}XRR \cite{Bertram:2007}. The magnetic asymmetry ratio $\Delta I=\frac{I_+-I_-}{I_++I_-}$ was simulated with ReMagX\cite{ReMagX} using the Zak matrix formalism \cite{Zak:1990a} and additional magnetooptic profiles for the change of optical constants with the position vertical to the film stack.
\begin{figure}[b!]
\centering
\includegraphics[width=8.5cm]{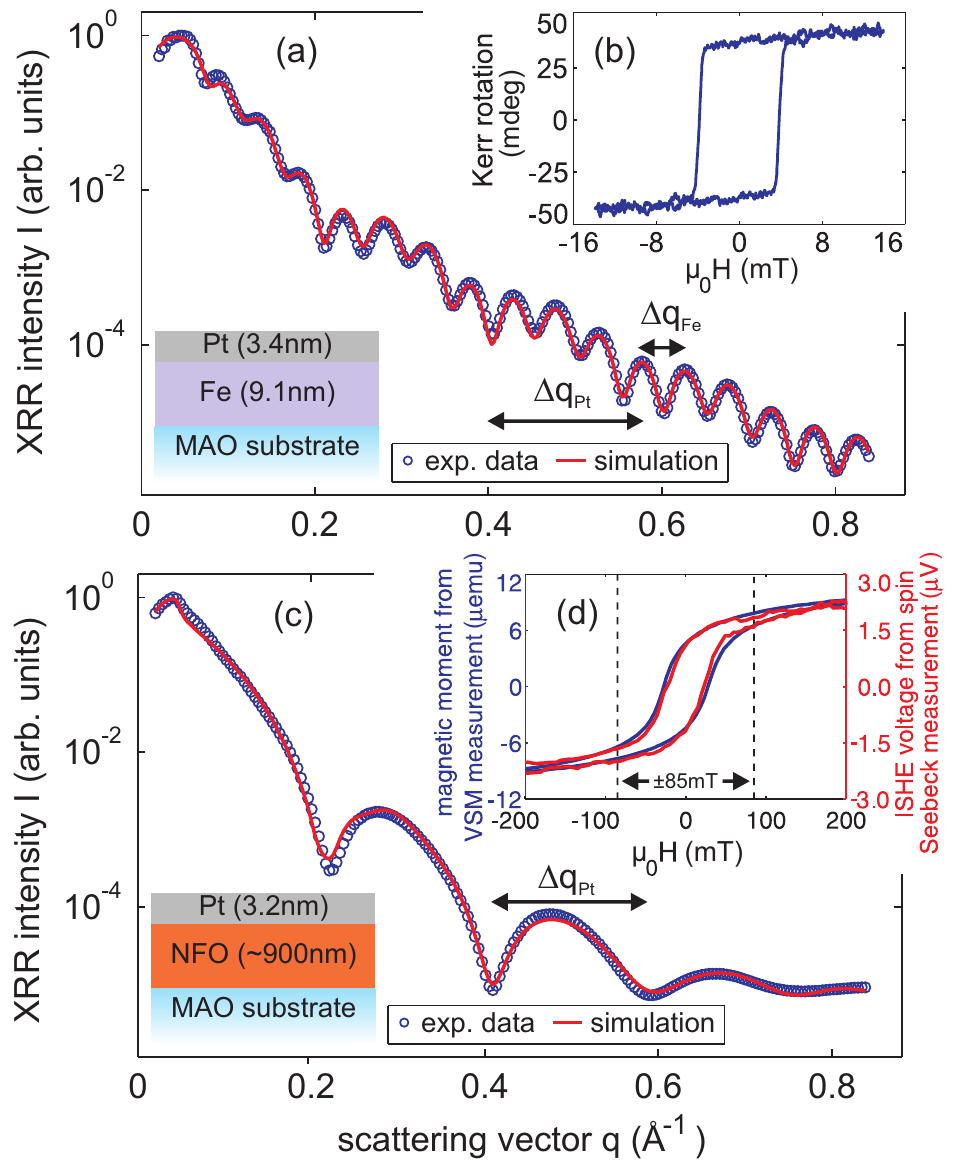}
\caption{(a)~XRR intensity $I$ and simulation for Pt/Fe/MAO. (b)~MOKE data ($\lambda=488\,\textrm{nm}$) for Pt/Fe/MAO. (c)~XRR intensity $I$ and simulation for Pt/NFO/MAO. (d)~Vibrating sample magnetometry (VSM) and LSSE  measurements ($\Delta T=18\,\textrm{K}$) for Pt/NFO/MAO.}
\label{fig:XRMR-XRRMOKE}
\end{figure}

To obtain the magnetic moment of the Pt, we calculated the absorption coefficients at the $L_3$ edge of Pt for various fixed spin moments with FDMNES \cite{Bunau:2009gp} in the full-multiple-scattering mode. The computed absorption spectra were fitted to experimental values far below and above the edge with the CHOOCH program \cite{Evans:2001fz}, which also produced the dispersion data by a Kramers-Kronig transformation. We checked that $\Delta\delta$ and $\Delta\beta$ are directly proportional to the magnetic moment of the Pt.

In Fig.~\ref{fig:XRMR-XRRMOKE}(a) the averaged XRR intensity $I(q)$ of Pt/Fe/MAO shows Kiessig fringes \cite{Kiessig:1931} with an oscillation length $\Delta q_{\textrm{Fe}}$ due to the interference of reflected light from the Pt/Fe and the Fe/MAO interfaces. The XRR analysis reveals a Fe layer thickness of $9.1\,\textrm{nm}$ with a roughness of $0.4\,\textrm{nm}$. The Pt film induces additional Kiessig fringes with the oscillation length $\Delta q_{\textrm{Pt}}$, which give a Pt thickness of $3.4\,\textrm{nm}$ and a roughness of $0.2\,\textrm{nm}$ for Pt and for the MAO substrate. During the experiment, the sample was saturated magnetically as deduced from magnetooptic Kerr effect (MOKE) measurements shown in Fig.~\ref{fig:XRMR-XRRMOKE}(b).

XRR from Pt/NFO/MAO (cf. Fig.~\ref{fig:XRMR-XRRMOKE}(c)) does not exhibit oscillations with $\Delta q_{\textrm{NFO}}$ due to the large thickness of the NFO. Therefore, the NFO acts as a quasisubstrate and the NFO/MAO interface is not accessible. The pronounced Kiessig fringes from the Pt layer, however, indicate a small roughness at both Pt interfaces. By fitting the intensity of the Kiessig fringes (cf. Fig.~\ref{fig:XRMR-XRRMOKE}(c)) the Pt thickness can be determined to $3.4\,\textrm{nm}$ while the roughness of the Pt/NFO and the air/Pt interfaces are $0.2\,\textrm{nm}$ and $0.4\,\textrm{nm}$, respectively. The roughness of the NFO/MAO interface has no appreciable influence on the reflectivity as shown and discussed in the supplemental material\cite{Kuschel:SM}. Since the NFO exhibits a very low MOKE, vibrating sample magnetometry (VSM) and LSSE measurements were performed (cf. Fig.~\ref{fig:XRMR-XRRMOKE}(d)). The results reveal a magnetic moment in the magnetic field direction of $70\%$ saturation for the magnetic field of $\pm85\,\textrm{mT}$ which is sufficient to observe dichroic effects.
\begin{figure}[t!]
\centering
\includegraphics[width=8.5cm]{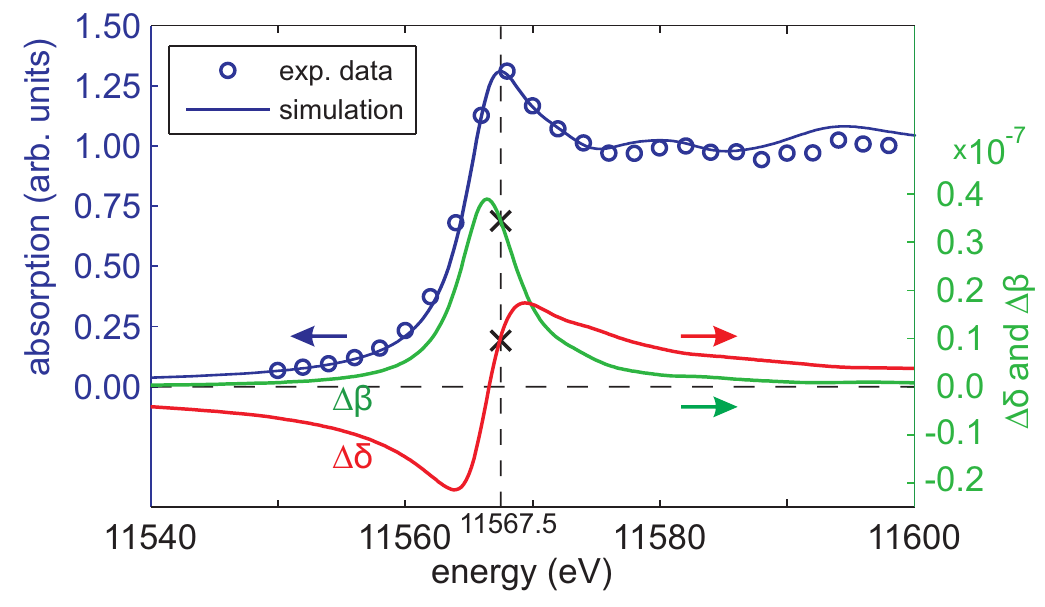}
\caption{Fluorescence spectrum at the Pt $L_3$ edge of the Pt($3.4\,\textrm{nm}$)/Fe($9.1\,\textrm{nm}$) sample and simulation, both normalized to the edge jump. $\Delta\delta$ and $\Delta\beta$ are derived from the simulation of the absorption resulting from a magnetic moment of $0.2\,\mu_{B}$ per Pt atom. XRMR was performed at the absorption maximum of $11567.5~\textrm{\textrm{eV}}$ (dashed vertical line) yielding a ratio $\Delta\beta/\Delta\delta$ of $3.4$ (black crosses).}
\label{fig:XRMR-Pt-Fe-edge}
\end{figure}

The absorption of the Pt($3.4\,\textrm{nm}$)/Fe($9.1\,\textrm{nm}$) sample at the Pt $L_3$ edge is presented in Fig.~\ref{fig:XRMR-Pt-Fe-edge}. The experimental data obtained by fluorescence fit closely to the simulation. The whiteline intensity (ratio of absorption maximum and edge jump) is 1.31, which indicates a mainly metallic state for Pt (1.25 for metallic Pt, 1.50 for PtO$_{1.36}$, 2.20 for PtO$_{1.6}$) \cite{Kolobov:2005eu}.

The asymmetry ratio of the Pt/Fe sample has its maximum near the absorption maximum of the Pt $L_3$ edge ($11567.5\,\textrm{eV}$) and vanishes at energies of more than $30\,\textrm{eV}$ below and above\cite{Klewe:2015}. Therefore, influences from other absorption edges can be excluded. Assuming a magnetic moment of $0.2\,\mu_{B}$ per Pt atom we calculated a maximum $\Delta\beta$ of $0.4\,\cdot\,10^{-7}$ which is located $1\,\textrm{eV}$ below the maximum absorption of Pt in accordance with the literature \cite{Schuetz:1990sd}. Since we performed XRMR measurements at the absorption maximum ($11567.5\,\textrm{eV}$, dashed vertical line in Fig.~\ref{fig:XRMR-Pt-Fe-edge}), we assume a ratio $\Delta\beta/\Delta\delta$ of $3.4$ (black crosses in Fig.~\ref{fig:XRMR-Pt-Fe-edge}) in the simulations of the asymmetry ratio.

\begin{figure}[t!]
\centering
\includegraphics[width=8.5cm]{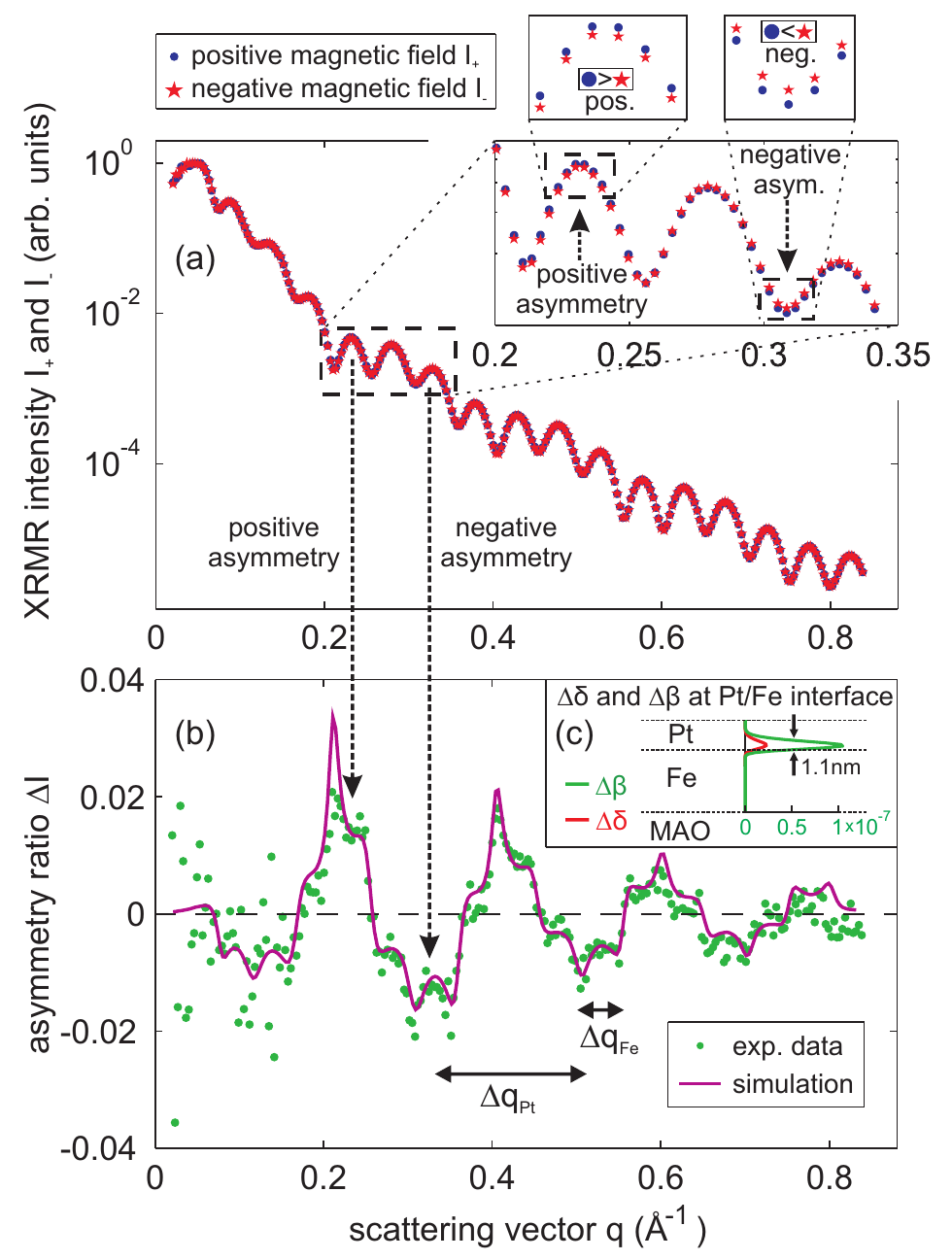}
\caption{(a) XRMR intensity $I_{\pm}$ for positive and negative external field and left circularly polarized light (photon energy of $11567.5\,\textrm{eV}$) for Pt($3.4\,\textrm{nm}$)/Fe($9.1\,\textrm{nm}$). (b)~Corresponding asymmetry ratio $\Delta I(q)$ and simulated data. (c)~Magnetooptic profile used for the asymmetry simulation.}
\label{fig:XRMR-Pt-Fe}
\end{figure}

Fig.~\ref{fig:XRMR-Pt-Fe}(a) presents the XRMR data at $11567.5~\textrm{eV}$ for Pt($3.4\,\textrm{nm}$)/Fe($9.1\,\textrm{nm}$). The blue dots and red stars denote the reflected intensity $I_{\pm}$ for positive and negative external magnetic field, respectively. An oscillation of $\Delta I (q)$ with an amplitude of about $2\%$ can be observed as shown in Fig.~\ref{fig:XRMR-Pt-Fe}(b). The asymmetry ratio can be fitted using the magnetooptic profile of $\Delta\delta$ and $\Delta\beta$ shown in the inset of Fig.~\ref{fig:XRMR-Pt-Fe}(b) while keeping the ratio $\Delta\beta/\Delta\delta=3.4$ constant. The thickness and roughness of the layers are kept fixed (taken from the XRR fit in Fig.~\ref{fig:XRMR-XRRMOKE}). The magnetooptic profile is generated by a Gaussian function at the Pt/Fe interface convoluted with the roughness profile of the layers\cite{Klewe:2015}. We obtain a FWHM value of the magnetooptic profile of $(1.1\pm0.1)\,\textrm{nm}$, which can be interpreted as the effective thickness of the spin polarized Pt layer at the interface to Fe.

By comparison of the experimental fit values of $\Delta\beta$ with the ab initio calculations we deduce a magnetic moment of $(0.6\pm0.1)\,\mu_{B}$ per Pt atom at the maximum of the magnetooptic profile. This value can also be found for spin polarized Pt in the literature \cite{Antel:1999,Wilhelm:2003di}. Compared to the results of Gepr\"ags et al. \cite{Gepraegs:2012be} we observe a larger magnetic moment for our reference sample. With XMCD, they detected an average magnetic moment of $0.0325\,\mu_{B}$ in Pt($10\,\textrm{nm}$)/Fe($10\,\textrm{nm}$). Assuming the spin polarized Pt is located in a $1.1\,\textrm{nm}$ layer, one obtains $0.0325\,\mu_{B}\cdot\frac{10\,\textrm{nm}}{1.1\,\textrm{nm}}\approx0.3\,\mu_{B}$ per Pt atom from these XMCD results.

\begin{figure}[t!]
\centering
\includegraphics[width=8.5cm]{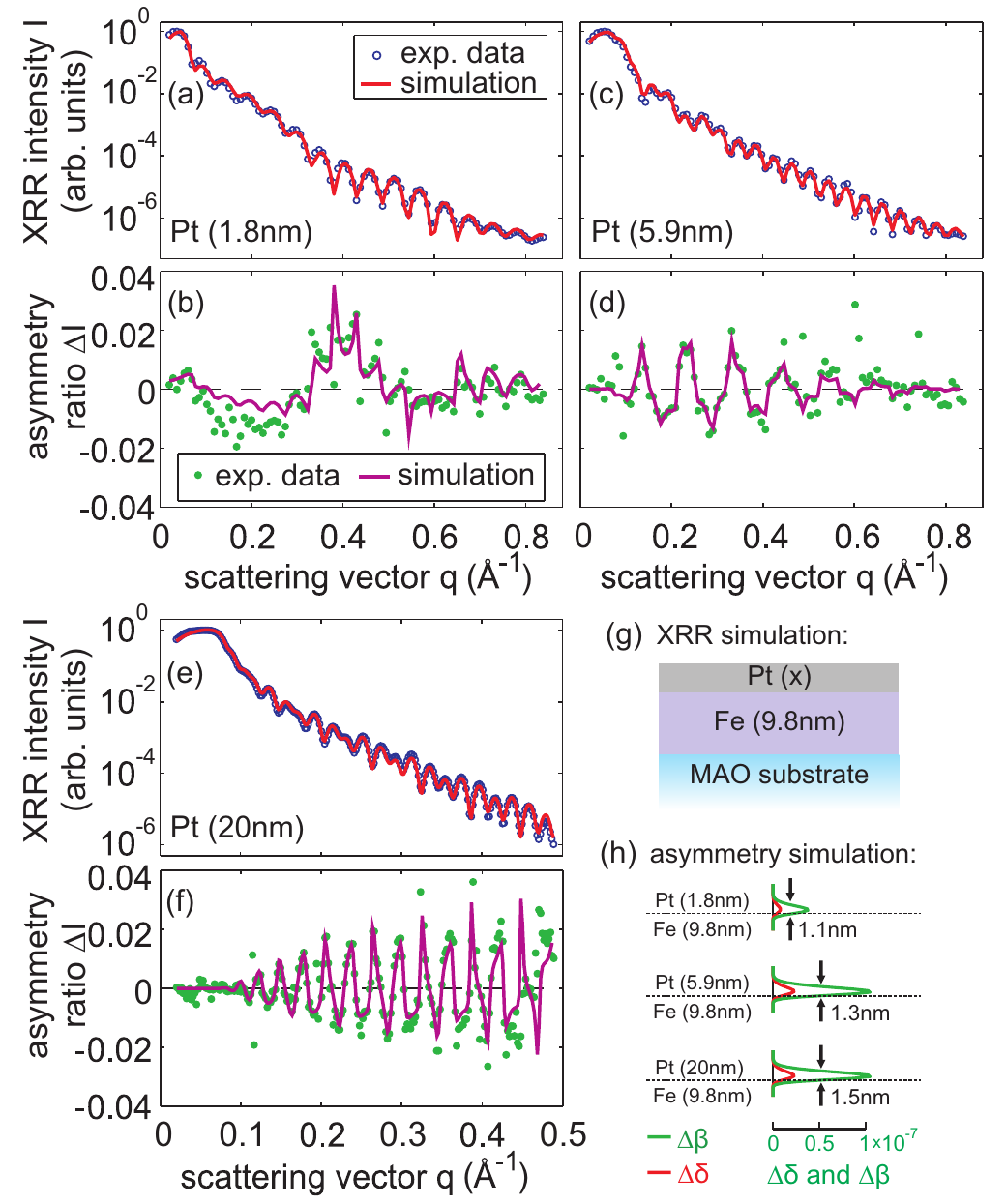}
\caption{XRR intensity $I$ and corresponding asymmetry ratios $\Delta I$ for Pt($x$)/Fe($9.8\,\textrm{nm}$) with (a),(b)~$x=1.8\,\textrm{nm}$ (average of eight curves), (c),(d)~$x=5.9\,\textrm{nm}$ and (e),(f)~$x=20\,\textrm{nm}$. For the XRR simulations the model in (g) is used and for the asymmetry simulations the magnetooptic profiles in (h) are used.}
\label{fig:XRMR-Pt-Fe-Thickness}
\end{figure}

We additionally investigated Pt($x$)/Fe($9.8\,\textrm{nm}$) for $x=1.8$, $5.9$ and $20$. The experimental XRR results and the simulations are presented in Fig.~\ref{fig:XRMR-Pt-Fe-Thickness}(a), \ref{fig:XRMR-Pt-Fe-Thickness}(c) and \ref{fig:XRMR-Pt-Fe-Thickness}(e). The corresponding asymmetry ratios are shown in Fig.~\ref{fig:XRMR-Pt-Fe-Thickness}(b), \ref{fig:XRMR-Pt-Fe-Thickness}(d) and \ref{fig:XRMR-Pt-Fe-Thickness}(f). For all Pt/Fe samples we observe an amplitude of about $2\%$, independent of the Pt thickness. This result shows the benefits of XRMR as the asymmetry signal stems from the spin polarization at the interface. The XRR model in Fig.~\ref{fig:XRMR-Pt-Fe-Thickness}(g) and the magnetooptic profiles in Fig.~\ref{fig:XRMR-Pt-Fe-Thickness}(h) are used for the simulations. 

We again obtain maximum magnetic moments of $(0.6\pm0.1)\,\mu_{B}$ for the sample with $5.9$ and $20\,\textrm{nm}$ Pt, while we observe a smaller maximum magnetic moment of $(0.2\pm0.1)\,\mu_{B}$ for the sample with $1.8\,\textrm{nm}$ Pt. This reduced moment might point to alteration of the thin Pt layer.

For Pt/NFO we obtain a whiteline intensity of 1.33 which is comparable to the value 1.31 for Pt/Fe and again confirms the mainly metallic Pt state in our samples. However, this value is slightly larger than the whiteline intensities of the Pt layers in the study of Gepr\"ags et al. \cite{Gepraegs:2012be, Geprags:2013} which were lower than 1.30, and it is lower than the value of the Pt layer in the study of Lu et al. \cite{Lu:2013eq}. Their whiteline intensity can be determined to be 1.45 \cite{Geprags:2013} which points to partly oxidized Pt when compared with 1.50 for PtO$_{1.36}$ \cite{Kolobov:2005eu}.

\begin{figure}[t!]
\centering
\includegraphics[width=8.5cm]{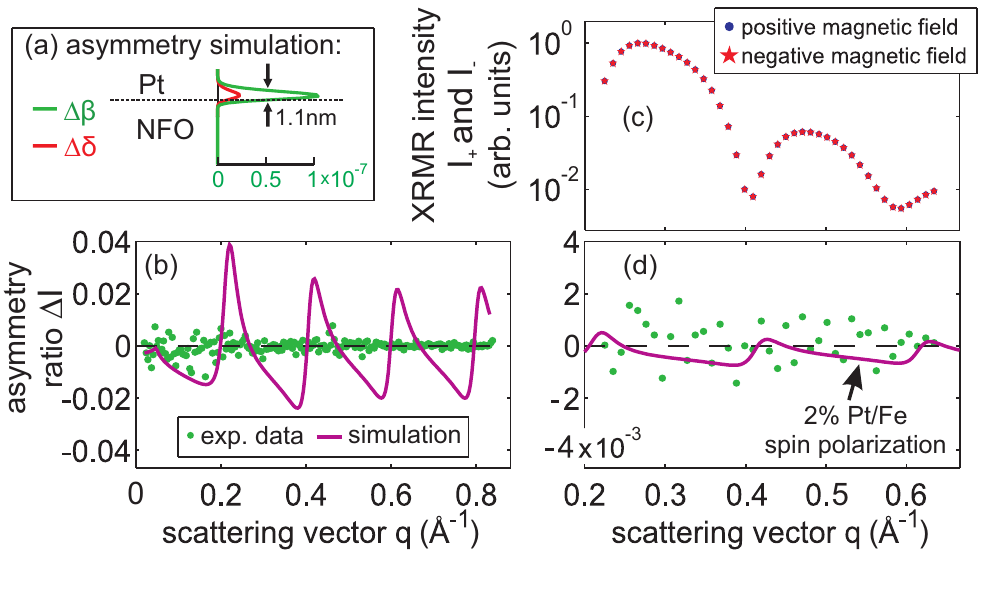}
\caption{XRMR intensity $I_{\pm}$ and corresponding asymmetry ratios $\Delta I(q)$ for Pt($3.2\,\textrm{nm}$)/ NFO($\sim$$900\,\textrm{nm}$) (cf. XRR curve in Fig.~\ref{fig:XRMR-XRRMOKE}(c)). (a) Magnetooptic profile used for the simulation in (b). (b)~Asymmetry ratio $\Delta I(q)$ (average of eight curves). (c)~XRMR intensity $I_{\pm}$ (average of 52 curves) for the reflectivity from $q=0.2\,$\AA$^{-1}$ up to $q=0.6\,$\AA$^{-1}$. (d)~Asymmetry ratio $\Delta I(q)$ from the XRMR in (c) and simulation assuming 2\% of the Pt/Fe spin polarization.}
\label{fig:XRMR-Pt-NFO}
\end{figure}

In Fig.~\ref{fig:XRMR-Pt-NFO}, the XRMR measurement of Pt/NFO is presented. For comparison, the measured asymmetry ratio $\Delta I(q)$ and a simulated asymmetry ratio are shown in Fig.~\ref{fig:XRMR-Pt-NFO}(b). Here, the magnetooptic profile of the spin polarization at the Pt/Fe interface  (cf. Fig.~\ref{fig:XRMR-Pt-NFO}(a)) was used for simulating a similar spin polarization in the Pt/NFO sample. Because of the different optical constants for NFO compared to Fe, the amplitude of $\Delta I(q)$ would be larger for Pt/NFO compared to Pt/Fe using the same magnetooptic profile. The oscillations obtained by this simulation are clearly not visible in the experimental data.

In order to increase the sensitivity, we measured the reflectivity from $q=0.2\,$\AA$^{-1}$ up to $q=0.6\,$\AA$^{-1}$ (two Kiessig fringes) and took the average from 52 experiments. The XRMR curves in Fig.~\ref{fig:XRMR-Pt-NFO}(c) still have a vanishing $\Delta I(q)$ (cf. Fig.~\ref{fig:XRMR-Pt-NFO}(d)) without any oscillations. A simulated asymmetry ratio assuming 2\% of the Pt/Fe spin polarization leads to the lower detection limit and can be converted into a maximum magnetic moment in Pt on NFO of $0.02\,\mu_{B}$ per Pt atom using the Pt/Fe calibration and taking into account that NFO is 70\% magnetized for $\pm85\,\textrm{mT}$ (cf. Fig.~\ref{fig:XRMR-XRRMOKE}(d)). Therefore, a static magnetic proximity effect in Pt/NFO can be neglected down to that limit.

In conclusion, we investigated the spin polarization of Pt in Pt/Fe and Pt/NFO bilayers using XRMR. We observed an asymmetry ratio of $2\%$ for the Pt/Fe samples independent of the Pt thickness. Simulations lead to a maximum magnetic moment of $(0.6\pm 0.1)\,\mu_{B}$ per Pt atom. The results demonstrate that XRMR is sensitive to the spin polarization at the Pt/FM interface independent from the Pt thickness. This feature together with the determination of the magnetooptic profiles for the change of optical constants and the ability to easily identify or exclude oscillating asymmetry ratios in noisy data endorse the use of XRMR in addition to XMCD for interface studies. For Pt/NFO samples no asymmetry was found. Therefore, a static magnetic proximity effect and proximity induced ANE can be excluded in recently reported LSSE investigations on Pt/NFO bilayers. The study of Pt spin polarization of Pt/YIG via XMCD and Pt/NFO via XRMR will be the start of future research on the interface spin polarization in NM/FMI bilayers. 

The authors gratefully acknowledge financial support by the Deutsche Forschungsgemeinschaft (DFG) within the priority program Spin Caloric Transport (SPP 1538) and the Deutsche Elektronen Synchrotron (DESY). They further acknowledge technical support by David Reuther at beamline P09 and thank Sebastian Macke for the software support of ReMagX. The work at the University of Alabama was supported by NSF-ECCS Grant No. 1102263.


\begin{thebibliography}{50}

\bibitem{Wolf:2001}
{S.~A.~Wolf, D.~D.~Awschalom, R.~A.~Buhrman, J.~M.~Daughton, S.~von~Moln\'a, M.~L.~Roukes, A.~J.~Chtchelkanova, D.~M.~Treger, Science~\textbf{294}, 1488 (2001).}

\bibitem{Bauer:2012fq}
{G.~E.~W.~Bauer, E.~Saitoh, B.~J.~van~Wees, Nature Mater.~\textbf{11}, 391 (2012),}

\bibitem{Tserkovnyak:2002jk}
{Y.~Tserkovnyak, A.~Brataas, G.~E.~W.~Bauer, Phys. Rev. Lett.~\textbf{88}, 117601 (2002).}

\bibitem{Hirsch:1999hn}
{J.~E.~Hirsch, Phys. Rev. Lett.~\textbf{83}, 1834 (1999).}

\bibitem{Uchida:2008cc}
{K.~Uchida, S.~Takahashi, K.~Harii, J.~Ieda, W.~Koshibae, K.~Ando, S.~Maekawa, E.~Saitoh, Nature~\textbf{455}, 778 (2008).}

\bibitem{Saitoh:2006kk}  
{E.~Saitoh, M.~Ueda, H.~Miyajima, G.~Tatara, Appl. Phys. Lett.~\textbf{88}, 182509 (2006).}

\bibitem{Hoffmann:2013el}
{A.~Hoffmann, IEEE Trans. Magn.~\textbf{49}, 5172 (2013).}

\bibitem{Huang:2012ks}
{S.~Y.~Huang, X.~Fan, D.~Qu, Y.~P.~Chen, W.~G.~Wang, J.~Wu, T.~Y.~Chen, J.~Q.~Xiao, C.~L.~Chien, Phys. Rev. Lett.~\textbf{109}, 107204 (2012).}

\bibitem{Ky:1966b}
{V.~D.~Ky, Phys. Status Solidi B~\textbf{17}, K207 (1966).}

\bibitem{Avery:2012bj}
{A.~D.~Avery, M.~R.~Pufall, B.~L.~Zink, Phys. Rev. Lett.~\textbf{109}, 196602, (2012).}
\
\bibitem{Schmid:2013ba}
{M.~Schmid, S.~Srichandan, D.~Meier, T.~Kuschel, J.-M.~Schmalhorst, M.~Vogel, G.~Reiss,  C.~Strunk, C.~H.~Back, Phys. Rev. Lett.~\textbf{111}, 187201 (2013).}

\bibitem{Meier:2013fa}
{D.~Meier, D.~Reinhardt, M.~Schmid, C.~H.~Back, J.-M.~Schmalhorst, T.~Kuschel, G.~Reiss, Phys. Rev. B~\textbf{88}, 184425 (2013).}

\bibitem{Nernst:1886}
{A.~von~Ettingshausen, W.~Nernst, Ann. Phys. Chem.~\textbf{265}, 343 (1886).}

\bibitem{Bosu:2011bw}
{S.~Bosu, Y.~Sakuraba, K.~Uchida, K.~Saito, T.~Ota, E.~Saitoh, K.~Takanashi, Phys. Rev. B~\textbf{83}, 224401 (2011).}

\bibitem{Huang:2011io}
{S.~Y.~Huang, W.~G.~Wang, S.~F.~Lee, J.~Kwo, C.~L.~Chien, Phys. Rev. Lett.~\textbf{107}, 216604 (2011).}

\bibitem{Meier2014}
{D.~Meier, D.~Reinhardt, M.~van~Straaten, C.~Klewe, M.~Althammer, M.~Schreier, S.~T.~B.~Goennenwein, A.~Gupta, M.~Schmid, C.~H.~Back, J.-M.~Schmalhorst, T.~Kuschel, G.~Reiss, Nat. Commun. \textbf{6}, 8211 (2014).}
 
\bibitem{Uchida:2010jb}
{K.~Uchida, H.~Adachi, T.~Ota, H.~Nakayama, S.~Maekawa, E.~Saitoh, Appl. Phys. Lett.~\textbf{97}, 172505 (2010).}

\bibitem{Meier:2013dz}
{D.~Meier, T.~Kuschel, L.~Shen, A.~Gupta, T.~Kikkawa, K.~Uchida, E.~Saitoh, J.-M.~Schmalhorst, G.~Reiss, Phys. Rev. B~\textbf{87}, 054421 (2013).}

%

\bibitem{Antel:1999}
{W.~J.~Antel, Jr., M.~M.~Schwickert, Tao~ Lin, W.~L.~O'Brien, G.~R.~Harp, Phys. Rev. B~\textbf{60}, 12933 (1999).}

\bibitem{Schuetz:1990sd}
{G.~Sch\"utz, R.~Wienke, W.~Wilhelm, W.~B.~Zeper, H.~Ebert, K.~Sp\"orl, J. Appl. Phys.~\textbf{67}, 4456 (1990).}

\bibitem{Wilhelm:2003di}
{F.~Wilhelm, P.~Poulopoulos, A.~Scherz, H.~Wende, K.~Baberschke, M.~Angelakeris, N.~K.~Flevaris, J.~Goulon, A.~Rogalev, Phys. Status Solidi A~\textbf{196}, 33 (2003).}

\bibitem{Wilhelm:2000}
{F.~Wilhelm, P.~Poulopoulos, G.~Ceballos, H.~Wende, K.~Baberschke, P.~Srivastava, D.~Benea, H.~Ebert, M.~Angelakeris, N.~K.~Flevaris, D.~Niarchos, A.~Rogalev, N.~B.~Brookes, Phys. Rev. Lett.~\textbf{85}, 413 (2000).}

\bibitem{Gepraegs:2012be}
{S.~Gepr\"ags, S.~Meyer, S.~Altmannshofer, M.~Opel, F.~Wilhelm, A.~Rogalev, R.~Gross, S.~T.~B.~Goennenwein, Appl. Phys. Lett.~\textbf{101}, 262407 (2012).}

\bibitem{Geprags:2013}
{S.~Gepr\"ags, S.~T.~B.~Goennenwein, M.~Schneider, F.~Wilhelm, K.~Ollefs, A.~Rogalev, M.~Opel, R.~Gross, Comment on Lu et al. (2013), arxiv:1307.4869.}

\bibitem{Lu:2013eq}
{Y.~M.~Lu, Y.~Choi, C.~M.~Ortega, X.~M.~Cheng, J.~W.~Cai, S.~Y.~Huang, L.~Sun, C.~L.~Chien, Phys. Rev. Lett.~\textbf{110}, 147207 (2013).}

\bibitem{Macke:2014em}
{S.~Macke, E.~Goering, J. Phys.: Condens. Matter~\textbf{26}, 363201 (2014).}

\bibitem{Althammer:2013ck}
{M.~Althammer, S.~Meyer, H.~Nakayama, M.~Schreier, S.~Altmannshofer, M.~Weiler, H.~Huebl, S.~Gepr\"ags, M.~Opel, R.~Gross, D.~Meier, C.~Klewe, T.~Kuschel, J.-M.~Schmalhorst, G.~Reiss, L.~Shen, A.~Gupta, Y.-T.~Chen, G.~E.~W.~Bauer, E.~Saitoh, S.~T.~B.~Goennenwein, Phys. Rev. B~\textbf{87}, 224401 (2013).}

\bibitem{Li:2011}
{N.~Li, Y.~H.~A.~Wang, M.~N.~Iliev, T.~M.~Klein, A.~Gupta, Chem. Vap. Deposition~\textbf{17}, 261 (2011).}

\bibitem{Strempfer:2013th}
{J.~Strempfer, S.~Francoual, D.~Reuther, D.~K.~Shukla, A.~Skaugen, H.~Schulte-Schrepping, T.~Kracht, H.~Franz, J. Synchrotron Rad.~\textbf{20}, 541 (2013).}

\bibitem{Parratt:1954fy}
{L.~G.~Parratt, Phys. Rev.~\textbf{95}, 359 (1954).}

\bibitem{Nevot:1980}
{L.~N\`evot, P.~Croce, Revue de Physique Appliquee~\textbf{15}, 761 (1980).}

\bibitem{Bertram:2007}
{F.~Bertram, Bachelor thesis, University of Osnabr\"uck (2007).}

\bibitem{ReMagX}
{www.remagx.org}

\bibitem{Zak:1990a}
{J.~Zak, E.~R.~Moog, C.~Liu, S.~D.~Bader, J. Magn. Magn. Mater.~\textbf{89}, 107 (1990).}

\bibitem{Bunau:2009gp}
{O.~Bun\v au, Y.~Joly, J. Phys.: Condens. Matter~\textbf{21}, 345501 (2009).}

\bibitem{Evans:2001fz}
{G.~Evans, R.~F.~Pettifer, J. Appl. Crystallogr.~\textbf{34}, 82 (2001).}

\bibitem{Kiessig:1931}
{H.~Kiessig, Ann. Phys.~\textbf{10}, 715 (1931).}

\bibitem{Kuschel:SM}
{Supplementary material, URL will be inserted by publisher.}

\bibitem{Kolobov:2005eu}
{A.~V.~Kolobov, F.~Wilhelm, A.~Rogalev, T.~Shima, J.~Tominaga, Appl. Phys. Lett.~\textbf{86}, 121909 (2005).}

\bibitem{Klewe:2015}
{C.~Klewe, T.~Kuschel, J.-M.~Schmalhorst, F.~Bertram, O.~Kuschel, J.~Wollschl\"ager, J.~Strempfer, M.~Meinert, G.~Reiss, arXiv:1508.00379 (2015).}

\end{thebibliography}
\end{document}